\documentclass{www2012-accepted}

\usepackage{latexsym}
\usepackage{amsmath}
\usepackage{graphicx}
\usepackage{multirow}
\usepackage[hyphens]{url}
\usepackage[tight,footnotesize]{subfigure}

\newcommand{\given}{\,|\,}

\usepackage{color}

\begin{document}

\clubpenalty=10000 
\widowpenalty = 10000

\title{Estimating the Prevalence of Deception\\ in Online Review Communities}

\numberofauthors{3}
\author{
\alignauthor Myle Ott\\
       \affaddr{Dept. of Computer Science}\\
       \affaddr{Cornell University}\\
       \affaddr{Ithaca, NY 14850}\\
       \email{myleott@cs.cornell.edu}
\alignauthor Claire Cardie\\
       \affaddr{Depts. of Computer Science and Information Science}\\
       \affaddr{Cornell University}\\
       \affaddr{Ithaca, NY 14850}\\
       \email{cardie@cs.cornell.edu}
\alignauthor Jeff Hancock\\
       \affaddr{Depts. of Communication and Information Science}\\
       \affaddr{Cornell University}\\
       \affaddr{Ithaca, NY 14850}\\
       \email{jeff.hancock@cornell.edu}
}

\maketitle

\begin{abstract}
Consumers' purchase decisions are increasingly influenced by user-generated online reviews~\cite{Cone:11}. Accordingly, there has been growing concern about the potential for posting \emph{deceptive opinion spam}---fictitious reviews that have been deliberately written to sound authentic, to deceive the reader~\cite{Ott:11}. But while this practice has received considerable public attention and concern, relatively little is known about the actual \emph{prevalence}, or rate, of deception in online review communities, and less still about the factors that influence it.

We propose a generative model of deception which, in conjunction with a deception classifier~\cite{Ott:11}, we use to explore the prevalence of deception in six popular online review communities: Expedia, Hotels.com, Orbitz, Priceline, TripAdvisor, and Yelp. We additionally propose a theoretical model of online reviews based on economic signaling theory~\cite{Spence:73}, in which consumer reviews diminish the inherent information asymmetry between consumers and producers, by acting as a signal to a product's true, unknown quality. We find that deceptive opinion spam is a growing problem overall, but with different growth rates across communities. These rates, we argue, are driven by the different signaling costs associated with deception for each review community, e.g., posting requirements. When measures are taken to increase signaling cost, e.g., filtering reviews written by first-time reviewers, deception prevalence is effectively reduced.

\end{abstract}

\category{I.2.7}{Artificial Intelligence}{Natural Language Processing}
\category{J.4}{Computer Applications}{Social and Behavioral Sciences}[economics, psychology]
\category{K.4.1}{Computers and Society}{Public Policy Issues}[abuse and crime involving computers]
\category{K.4.4}{Computers and Society}{Electronic Commerce}

\terms{Algorithms, Experimentation, Measurement, Theory}

\keywords{Deceptive opinion spam, Deception prevalence, Gibbs sampling, Online reviews, Signaling theory}

\section{Introduction} \label{sec:intro}

Consumers rely increasingly on user-generated online reviews to make,
or reverse, purchase decisions~\cite{Cone:11}. Accordingly, there
appears to be widespread and growing concern among both
businesses and the public~\cite{CNET:09,NYTimes:09,Guardian:11,NYTimes:11,NYTimes:12,Guardian:10}
regarding the potential for posting \emph{deceptive opinion
  spam}---fictitious reviews that have been deliberately written to sound authentic, to deceive the
reader~\cite{Ott:11}. Perhaps surprisingly, however, relatively little is known about the
actual \emph{prevalence}, or rate, of deception in online
review communities, and less still is known about the factors that can
influence it.  On the one hand, the relative ease of producing reviews, 
combined with the pressure for businesses, products, and services
to be perceived in a positive light, might lead one to expect that 
a preponderance of online reviews are fake.
One can argue, on the other hand, that a low rate of deception is
required for review sites to serve any value.\footnote{It is worth pointing out that a review site containing deceptive reviews might still serve value, for example, if there remains enough truthful content to produce reasonable aggregate comparisons between offerings.}

The focus of spam research in the context of online reviews has been primarily on \emph{detection}. Jindal and Liu~\cite{Jindal:08}, for example, train models using features based on the review text, reviewer, and product to identify \emph{duplicate} opinions.\footnote{Duplicate (or near-duplicate) opinions are opinions that appear more than once in the corpus with the same (or similar) text. However, simply because a review is duplicated does not make it deceptive. Furthermore, it seems unlikely that either duplication or plagiarism characterizes the majority of fake reviews. Moreover, such reviews are potentially detectable via off-the-shelf plagiarism detection software.} Yoo and Gretzel~\cite{Yoo:09} gather 40 truthful and 42 deceptive hotel reviews and, using a standard statistical test, manually compare the psychologically relevant linguistic differences between them. While useful, these approaches do not focus on the prevalence of deception in online reviews.

Indeed, empirical, scholarly studies of the prevalence
of deceptive opinion spam have remained elusive. One reason is the
difficulty in obtaining reliable \emph{gold-standard
annotations} for reviews, i.e., trusted labels that tag each review as either
truthful (real) or deceptive (fake).
One option for producing gold-standard labels, for example, would be
to rely on the judgements of human annotators.  Recent studies, however, show that
deceptive opinion spam is not easily identified by human
readers~\cite{Ott:11}; this is especially the case when considering
the overtrusting nature of most human judges, a phenomenon referred to
in the psychological deception literature as a \emph{truth
 bias}~\cite{Vrij:08}. To help illustrate the non-trivial nature of
identifying deceptive content, given below are two positive reviews of
the Hilton Chicago Hotel, one of which is truthful, and the other of
which is deceptive opinion spam:

\begin{enumerate}

\item {``My husband and I stayed in the Hilton Chicago and had a very
  nice stay! The rooms were large and comfortable. The view of Lake
  Michigan from our room was gorgeous. Room service was really good
  and quick, eating in the room looking at that view, awesome! The
  pool was really nice but we didnt get a chance to use it. Great
  location for all of the downtown Chicago attractions such as
  theaters and museums. Very friendly staff and knowledgable, you cant
  go wrong staying here."}

\item {``We loved the hotel. When I see other posts about it being
  shabby I can't for the life of me figure out what they are talking
  about. Rooms were large with TWO bathrooms, lobby was fabulous, pool
  was large with two hot tubs and huge gym, staff was courteous. For
  us, the location was great--across the street from Grant Park with a
  great view of Buckingham Fountain and close to all the museums and
  theatres. I'm sure others would rather be north of the river closer
  to the Magnificent Mile but we enjoyed the quieter and more scenic
  location. Got it for \$105 on Hotwire. What a bargain for such a
  nice hotel."}

\item[]{\small{Answer: See footnote.}\footnote{The first review is deceptive opinion
  spam.}}
\end{enumerate}  

The difficulty of detecting which of these reviews is fake is consistent with recent large meta-analyses demonstrating the inaccuracy of human judgments of deception, with accuracy rates typically near chance~\cite{Bond:06}. In particular, humans have a difficult time identifying deceptive messages from cues alone, and as such, it is not surprising that research on estimating the prevalence of deception (see Section~\ref{sec:implications}) has generally relied on self-report methods, even though such reports are difficult and expensive to obtain, especially in large-scale settings, e.g., the web~\cite{Hancock:09:DMCC}. More importantly, self-report methods, such as diaries and large-scale surveys, have several methodological concerns, including social desirability bias and self-deception~\cite{DePaulo:96}. Furthermore, there are considerable disincentives to revealing one's own deception in the case of online reviews, such as being permanently banned from a review portal, or harming a business's reputation.

Recently, automated approaches (see
Section~\ref{sec:dec_class}) have emerged to reliably label reviews as truthful
vs.\ deceptive: Ott et al.~\shortcite{Ott:11} train an $n$-gram--based
text classifier using a corpus of truthful and deceptive reviews---the former culled from online
review communities and the latter generated using Amazon Mechanical
Turk (\url{http://www.mturk.com}).  Their resulting classifier
is nearly 90\% accurate.

\emph{In this work, we present a general framework (see
  Section~\ref{sec:frame}) for estimating the prevalence of deception
  in online review communities.}  Given a classifier that
distinguishes truthful from deceptive reviews (like that described
above), and inspired by studies of disease
prevalence~\cite{Johnson:01,Joseph:95}, we propose a \emph{generative
  model of deception} (see Section~\ref{sec:prev_model}) that jointly
models the classifier's uncertainty as well as the ground-truth
deceptiveness of each review. Inference for this model, which we
perform via Gibbs sampling, allows us to estimate the prevalence of
deception in the underlying review community, \emph{without relying on either
  self-reports or gold-standard annotations}.

\emph{We further propose a theoretical component to the framework
based on signaling theory from economics~\cite{Spence:73} (see
Section~\ref{sec:signal_theory}) and use it to reason about the factors that 
influence deception prevalence in online review communities.}
In our context, signaling theory
interprets each review as a signal to the product's true, unknown quality;
thus, the goal of consumer reviews is to diminish the inherent information asymmetry between 
consumers and producer. Very briefly, according to a signaling theory approach, deception
prevalence should be a function of the costs and benefits that accrue from producing a fake review.
We hypothesize that review communities with low signaling cost, such as 
communities that make it easy to post a review, and large benefits, such as highly trafficked sites,
will exhibit more deceptive opinion spam than those with higher signaling costs,
such as communities that establish additional requirements for posting reviews, and lower
benefits, such as low site traffic.

We apply our approach to the domain of hotel reviews.
In particular, we examine
hotels from the Chicago area, restricting attention to \textbf{positive} reviews only,
and instantiate the framework for six
online review communities (see Section~\ref{sec:data}): Expedia
(\url{http://www.expedia.com}), Hotels.com
(\url{http://www.hotels.com}), Orbitz (\url{http://www.orbitz.com}),
Priceline (\url{http://www.priceline.com}), TripAdvisor
(\url{http://www.tripadvisor.com}), and Yelp
(\url{http://www.yelp.com}).

We find first that the prevalence of deception indeed varies by community.
However, because it is not possible to validate these estimates empirically
(i.e., the gold-standard rate of deception in each community is unknown), we focus our
discussion instead on the relative differences in the rate of deception between
communities. Here, the results confirm our hypotheses and suggest
that deception is most prevalent in 
communities with a low signal cost. Importantly, when measures are taken 
to increase a community's signal cost,
we find dramatic reductions in our estimates of the rate of deception
in that community.

\section{Framework} \label{sec:frame}

\emph{In this section, we propose a framework to estimate the prevalence, or rate, of deception among reviews in six online review communities.} Since reviews in these communities do not have gold-standard annotations of deceptiveness, and neither human judgements nor self-reports of deception are reliable in this setting (see discussion in Section~\ref{sec:intro}), our framework instead estimates the rates of deception in these communities using the output of an imperfect, automated deception classifier. In particular, we utilize a supervised machine learning classifier, which has been shown recently by Ott et al.~\cite{Ott:11} to be nearly 90\% accurate at detecting deceptive opinion spam in a class-balanced dataset.

A similar framework has been used previously in studies of disease prevalence, in which gold-standard diagnostic testing is either too expensive, or impossible to perform~\cite{Johnson:01,Joseph:95}. In such cases, it is therefore necessary to \emph{estimate} the prevalence of disease in the population using a combination of an imperfect diagnostic test, and estimates of the test's positive and negative recall rates.\footnote{Recall rates of an imperfect diagnostic test are unlikely to be known precisely. However, imprecise estimates can often be obtained, especially in cases where it is feasible to perform gold-standard testing on a small subpopulation.}

Our proposed framework is summarized here, with each step discussed in greater detail in the corresponding section:
\begin{enumerate}
\item \textbf{Data (Section~\ref{sec:data}):}

Assume given a set of labeled training reviews, $\mathcal{D}^{\text{train}} = \{(\mathbf{x}_i, y_i)\}_{i=1}^{N^{\text{train}}}$, where, for each review $i$, $y_i \in \{0, 1\}$ gives the review's label ($0$ for truthful, $1$ for deceptive), and $\mathbf{x}_i \in \mathbb{R}^{|V|}$ gives the review's feature vector representation, for some feature space of size $|V|$. Similarly, assume given a set of \emph{labeled truthful} development reviews, $\mathcal{D}^{\text{dev}} = \{(\mathbf{x}_i, 0)\}_{i=1}^{N^{\text{dev}}}$, and a set of \emph{unlabeled} test reviews, $\mathcal{D}^{\text{test}} = \{\mathbf{x}_i\}_{i=1}^{N^{\text{test}}}$.

\item \textbf{Deception Classifier (Section~\ref{sec:dec_class}):}

Using the labeled training reviews, $\mathcal{D}^{\text{train}}$, learn a supervised deception classifier, $f : \mathbb{R}^{|V|} \rightarrow \{0, 1\}$.

\item \textbf{Classifier Sensitivity and Specificity (Section~\ref{sec:class_sens_and_spec}):}

By cross-validation on $\mathcal{D}^{\text{train}}$, estimate the \emph{sensitivity} (deceptive recall) of the deception classifier, $f$, as:
  \begin{align}
  \eta = \Pr(f(\mathbf{x}_i) = 1 \given y_i = 1).
  \end{align}

Then, use $\mathcal{D}^{\text{dev}}$ to estimate the \emph{specificity} (truthful recall) of the deception classifier, $f$, as:
  \begin{align}
  \theta = \Pr(f(\mathbf{x}_i) = 0 \given y_i = 0).
  \end{align}

\item \textbf{Prevalence Models (Section~\ref{sec:prev_model}):}

Finally, use $f$, $\eta$, $\theta$, and either the Na\"ive Prevalence Model (Section~\ref{sec:prev_model_naive}), or the generative Bayesian Prevalence Model (Section~\ref{sec:prev_model_bayes}), to estimate the prevalence of deception, denoted $\pi$, among reviews in $\mathcal{D}^{\text{test}}$. Note that \emph{if we had gold-standard labels}, $\{y_i\}_{i=1}^{N^{\text{test}}}$, the gold-standard prevalence of deception would be:
  \begin{align}
  \pi^* = \frac{1}{N^{\text{test}}} \sum_{i = 1}^{N^{\text{test}}} y_i.
  \end{align}
\end{enumerate}

\section{Prevalence Models} \label{sec:prev_model}

In Section~\ref{sec:frame}, we propose a framework to estimate the prevalence of deception in a group of reviews using only the output of a noisy deception classifier. Central to this framework is the Prevalence Model, which models the uncertainty of the deception classifier, and ultimately produces the desired prevalence estimate. In this section, we propose two competing Prevalence Models, which can be used interchangeably in our framework.

\subsection{Na\"ive Prevalence Model} \label{sec:prev_model_naive}

The Na\"ive Prevalence Model (\textsc{na\"ive}) estimates the prevalence of deception in a corpus of reviews by correcting the output of a noisy deception classifier according to the classifier's \emph{known} performance characteristics.

Formally, for a given deception classifier, $f$, let $\pi_f$ be the number of reviews in $\mathcal{D}^{\text{test}}$ for which $f$ makes a positive prediction, i.e., the number of reviews for which $f$ predicts \emph{deceptive}. Also, let the sensitivity (deceptive recall) and specificity (truthful recall) of $f$ be given by $\eta$ and $\theta$, respectively. Then, we can write the expectation of $\pi_f$ as:
\begin{align}
\mathbb{E}[\pi_f] &= \mathbb{E}\left[ \frac{1}{N^{\text{test}}} \sum_{\mathbf{x} \in \mathcal{D}^{\text{test}}} \delta[f(\mathbf{x}) = 1] \right] \nonumber\\
  &= \frac{1}{N^{\text{test}}} \sum_{\mathbf{x} \in \mathcal{D}^{\text{test}}} \mathbb{E}\left[ \delta\left[f(\mathbf{x}) = 1\right] \right] \nonumber\\
  &= \eta \pi^* + (1 - \theta)(1 - \pi^*), \label{eqn:expect_f}
\end{align}
where $\pi^*$ is the true (latent) rate of deception, and $\delta[a = b]$ is the Kronecker delta function, which is equal to $1$ when $a = b$, and $0$ otherwise.

If we rearrange Equation~\ref{eqn:expect_f} in terms of $\pi^*$, and replace the expectation of $\pi_f$ with the observed value, we get the Na\"ive Prevalence Model estimator:
\begin{align}
\pi_{\textsc{na\"ive}} = \frac{\pi_f - (1 - \theta)}{\eta - (1 - \theta)}. \label{eqn:naive_prev_est}
\end{align}
Intuitively, Equation~\ref{eqn:naive_prev_est} corrects the raw classifier output, given by $\pi_f$, by subtracting from it the false positive rate, given by $1 - \theta$, and dividing the result by the difference between the true and false positive rates, given by $\eta - (1 - \theta)$. Notice that when $f$ is an oracle,\footnote{An oracle is a classifier that does not make mistakes, and always predicts the true, gold-standard label.} i.e., when $\eta = \theta = 1$, the Na\"ive Prevalence Model estimate correctly reduces to the oracle rate given by $f$, i.e., $\pi_{\textsc{na\"ive}} = \pi_f = \pi^*$.

\subsection{Bayesian Prevalence Model} \label{sec:prev_model_bayes}

\begin{figure}[t]
\begin{center}
\includegraphics[scale=0.65]{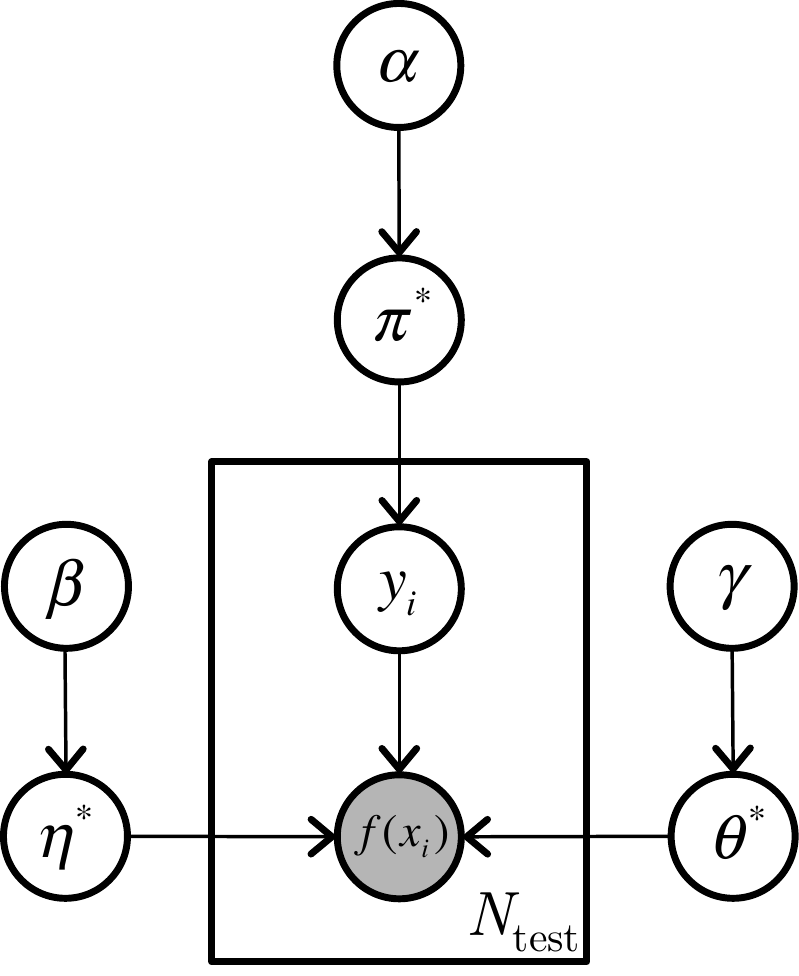}
\caption{The Bayesian Prevalence Model in plate notation. Shaded nodes represent observed variables, and arrows denote dependence. For example, $f(\mathbf{x}_i)$ is observed, and depends on $\eta^*$, $\theta^*$, and $y_i$.}
\label{fig:plate1}
\end{center}
\end{figure}

Unfortunately, the Na\"ive Prevalence Model estimate, $\pi_{\textsc{na\"ive}}$, is not restricted to the range $[0, 1]$. Specifically, it is negative when $\pi_f < 1 - \theta$, and greater than $1$ when $\pi_f > \eta$. Furthermore, the Na\"ive Prevalence Model makes the unrealistic assumption that the estimates of the classifier's sensitivity ($\eta$) and specificity ($\theta$), obtained using the procedure discussed in Section~\ref{sec:class_sens_and_spec} and Appendix~\ref{app:class_sens_and_spec}, are exact.

The Bayesian Prevalence Model (\textsc{bayes}) addresses these limitations by modeling the generative process through which deception occurs, or, equivalently, the joint probability distribution of the observed and latent data. In particular, \textsc{bayes} models the observed classifier output, the true (latent) rate of deception ($\pi^*$), as well as the classifier's true (latent) sensitivity ($\eta^*$) and specificity ($\theta^*$). Formally, \textsc{bayes} assumes that our data was generated according to the following generative story:
\begin{itemize}
\item Sample the true rate of deception: $\pi^* \sim \text{Beta}(\boldsymbol{\alpha})$
\item Sample the classifier's true sensitivity: $\eta^* \sim \text{Beta}(\boldsymbol{\beta})$
\item Sample the classifier's true specificity: $\theta^* \sim \text{Beta}(\boldsymbol{\gamma})$
\item For each review $i$: \begin{itemize}
  \item Sample the ground-truth deception label: \[ y_i \sim \text{Bernoulli}(\pi^*) \]
  \item Sample the classifier's output: \[
    f(\mathbf{x}_i) \sim \left\{ \begin{array}{l l}
      \text{Bernoulli}(\eta^*) & \text{if $y_i = 1$} \\
      \text{Bernoulli}(1 - \theta^*) & \text{if $y_i = 0$} \end{array} \right. \]
  \end{itemize}
\end{itemize}

The corresponding graphical model is given in plate notation in Figure~\ref{fig:plate1}. Notice that by placing Beta prior distributions on $\pi^*$, $\eta^*$, and $\theta^*$, \textsc{bayes} enables us to encode our prior knowledge about the true rate of deception, as well as our uncertainty about the estimates of the classifier's sensitivity and specificity. This is discussed further in Section~\ref{sec:class_sens_and_spec}.

A similar model has been proposed by Joseph et al.~\cite{Joseph:95} for studies of disease prevalence, in which it is necessary to estimate the prevalence of disease in a population given only an imperfect diagnostic test. However, that model samples the \emph{total number of true positives and false negatives}, while our model samples the $y_i$ individually. Accordingly, while pilot experiments confirm that the two models produce identical results, the generative story of our model, given above, is comparatively much more intuitive.

\subsubsection{Inference}

While exact inference is intractable for the Bayesian Prevalence Model, a popular alternative way of approximating the desired posterior distribution is with Markov Chain Monte Carlo (MCMC) sampling, and more specifically Gibbs sampling. Gibbs sampling works by sampling each variable, in turn, from the conditional distribution of that variable given all other variables in the model. After repeating this procedure for a fixed number of iterations, the desired posterior distribution can be approximated from samples in the chain by: (1) discarding a number of initial \emph{burn-in} iterations, and (2) since adjacent samples in the chain are often highly correlated, thinning the number of remaining samples according to a \emph{sampling lag}.

The conditional distributions of each variable given the others can be derived from the joint distribution, which can be read directly from the graph. Based on the graphical representation of \textsc{bayes}, given in Figure~\ref{fig:plate1}, the joint distribution of the observed and latent variables is just:
\begin{multline}
\Pr(f(\mathbf{x}), \mathbf{y}, \pi^*, \eta^*, \theta^*; \boldsymbol{\alpha}, \boldsymbol{\beta}, \boldsymbol{\gamma}) = \Pr(f(\mathbf{x}) \given \mathbf{y}, \eta^*, \theta^*) \cdot \\
\Pr(\mathbf{y} \given \pi^*) \cdot \Pr(\pi^* \given \boldsymbol{\alpha}) \cdot \Pr(\eta^* \given \boldsymbol{\beta}) \cdot \Pr(\theta^* \given \boldsymbol{\gamma}), \label{eqn:full_joint}
\end{multline}
where each term is given according to the sampling distributions specified in the generative story in Section~\ref{sec:prev_model_bayes}.

A common technique to simplify the joint distribution, and the sampling process, is to integrate out (collapse) variables that do not need to be sampled. If we integrate out $\pi^*$, $\eta^*$, and $\theta^*$ from Equation~\ref{eqn:full_joint}, we can derive a Gibbs sampler that only needs to sample the $y_i$'s at each iteration. The resulting sampling equations, and the corresponding Bayesian Prevalence Model estimate of the prevalence of deception, $\pi_{\textsc{bayes}}$, are given in greater detail in Appendix~\ref{app:bayes_inference}.

\section{Deception Detection} \label{sec:dec_det}

\subsection{Deception Classifier} \label{sec:dec_class}

\begin{table}[t]
\caption{Reference 5-fold cross-validated performance of an SVM deception detection classifier in a balanced dataset of TripAdvisor reviews, given by Ott et al.~\cite{Ott:11}. F-score corresponds to the harmonic mean of precision and recall.}
\label{table:dec_det_perf}
\begin{center}
\begin{tabular}{l|c}
\hline
\multicolumn{1}{c|}{\textsc{metric}} & \textsc{performance} \\
\hline
Accuracy & 89.6\% \\
\hline
Deceptive Precision & 89.1\% \\
Deceptive Recall & 90.3\% \\
Deceptive F-score & 89.7\% \\
\hline
Truthful Precision & 90.1\% \\
Truthful Recall & 89.0\% \\
Truthful F-score & 89.6\% \\
\hline \hline
Baseline Accuracy & 50\% \\
\hline
\end{tabular}
\end{center}
\end{table}

The next component of the framework given in Section~\ref{sec:frame} is the \emph{deception classifier}, which predicts whether each unlabeled review is truthful (real) or deceptive (fake). Following previous work~\cite{Ott:11}, we assume given some amount of \emph{labeled} training reviews, so that we can train deception classifiers using a \emph{supervised} learning algorithm.

Previous work has shown that Support Vector Machines (SVM) trained on $n$-gram features perform well in deception detection tasks~\cite{Jindal:08,Mihalcea:09,Ott:11}. Following Ott et al.~\cite{Ott:11}, we train \emph{linear} SVM classifiers using the LIBSVM~\cite{CC01a} software package, and represent reviews using unigram and bigram bag-of-words features. While more sophisticated and purpose-built classifiers might achieve better performance, pilot experiments suggest that the Prevalence Models (see Section~\ref{sec:prev_model}) are not heavily affected by minor differences in classifier performance. Furthermore, the simple approach just outlined has been previously evaluated to be nearly 90\% accurate at detecting deception in a balanced dataset~\cite{Ott:11}. Reference cross-validated classifier performance appears in Table~\ref{table:dec_det_perf}.

\subsection{Classifier Sensitivity and Specificity} \label{sec:class_sens_and_spec}

Both Prevalence Models introduced in Section~\ref{sec:prev_model} can utilize knowledge of the underlying deception classifier's \emph{sensitivity} ($\eta^*$), i.e., deceptive recall rate, and \emph{specificity} ($\theta^*$), i.e., truthful recall rate. While it is not possible to obtain gold-standard values for these parameters, we can obtain rough estimates of their values (denoted $\eta$ and $\theta$, respectively) through a combination of cross-validation, and evaluation on a labeled development set. For the Na\"ive Prevalence Model, the estimates are used directly, and are assumed to be exact.

For the Bayesian Prevalence Model, we adopt an empirical Bayesian approach and use the estimates to inform the corresponding Beta priors via their hyperparameters, $\boldsymbol{\beta}$ and $\boldsymbol{\gamma}$, respectively. The full procedure is given in Appendix~\ref{app:class_sens_and_spec}.

\section{Data} \label{sec:data}

\begin{table}[t]
\caption{Corpus statistics for unlabeled test reviews from six online review communities.}
\label{table:corp_stats}
\begin{center}
\begin{tabular}{l|cc}
\hline
\multicolumn{1}{c|}{\textsc{community}} & \textsc{\# hotels} & \textsc{\# reviews} \\
\hline
\textbf{Expedia} & 100 & 4,341 \\
\textbf{Hotels.com} & 103 & 6,792 \\
\textbf{Orbitz} & 97 & 1,777 \\
\textbf{Priceline} & 98 & 4,027 \\
\textbf{TripAdvisor} & 104 & 9,602 \\
\textbf{Yelp} & 103 & 1,537 \\
\hline \hline
\textbf{Mechanical Turk} & 20 & 400 \\
\hline
\end{tabular}
\end{center}
\end{table}

In this section, we briefly discuss each of the three kinds of data used by our framework introduced in Section~\ref{sec:frame}. Corpus statistics are given in Table~\ref{table:corp_stats}.
Following Ott et al.~\cite{Ott:11}, we excluded all reviews with fewer than 150 characters, as well as all non-English reviews.\footnote{Language was identified by the Language Detection Library: \url{http://code.google.com/p/language-detection/}.}

\subsection{Training Reviews ($\mathcal{D}^{\text{train}}$)}

Training a supervised deception classifier requires labeled training data. Following Ott et al.~\cite{Ott:11}, we build a balanced set of 800 training reviews,
containing 400 truthful reviews from six online review communities, and 400 gold-standard deceptive reviews from Amazon Mechanical Turk.

\emph{Deceptive Reviews}: In Section~\ref{sec:intro}, we discuss some of the difficulties associated with obtaining gold-standard labels of deception, including the inaccuracy of human judgements, and the problems with self-reports of deception. To avoid these difficulties, Ott et al.~\cite{Ott:11} have recently \emph{created} 400 gold-standard deceptive reviews using Amazon's Mechanical Turk service. In particular, they paid one US dollar (\$1) to each of 400 unique Mechanical Turk workers to write a fake \emph{positive} (5-star) review for one of the 20 most \emph{heavily-reviewed} Chicago hotels on TripAdvisor. Each worker was given a link to the hotel's website, and instructed to write a convincing review from the perspective of a satisfied customer. Any submission found to be plagiarized was rejected. Any submission with fewer than 150 characters was discarded. To date, this is the only publicly-available\footnote{\url{http://www.cs.cornell.edu/~myleott/op_spam}} gold-standard deceptive opinion spam dataset. As such, we choose it to be our sole source of labeled deceptive reviews for training our supervised deception classifiers. Note that these same reviews are used to estimate the resulting classifier sensitivity (deceptive recall), via the cross-validation procedure given in Appendix~\ref{app:class_sens_and_spec}.

\emph{Truthful Reviews}: Many of the same challenges that make it difficult to obtain gold-standard deceptive reviews, also apply to obtaining truthful reviews. Related work~\cite{Jindal:08,Lim:10} has hypothesized that the relative impact of spam reviews is smaller for heavily-reviewed products, and that therefore spam should be less common among them. For consistency with our labeled deceptive review data, we simply label as truthful all positive (5-star) reviews of the 20 previously chosen Chicago hotels. We then draw a random sample of size 400, and take that to be our labeled truthful training data.

\subsection{Development Reviews ($\mathcal{D}^{\text{dev}}$)}

By training on deceptive and truthful reviews from the same 20 hotels, we are effectively controlling our classifier for topic. However, because this training data is not representative of Chicago hotel reviews in general, it is important that we do not use it to estimate the resulting classifier's specificity (truthful recall). Accordingly, as specified in our framework (Section~\ref{sec:frame}), classifier specificity is instead estimated on a separate, \emph{labeled truthful} development set, which we draw uniformly at random from the unlabeled reviews in each review community. For consistency with the sensitivity estimate, the size of the draw is always 400 reviews.

\subsection{Test Reviews ($\mathcal{D}^{\text{test}}$)}

The last data component of our framework is the set of test reviews, among which to estimate the prevalence of deception. To avoid evaluating reviews that are too different from our training data in either sentiment (due to negative reviews), or topic (due to reviews of hotels outside Chicago), we constrain each community's test set to contain only \emph{positive (5-star) Chicago hotel reviews}. This unfortunately disqualifies our estimates of each community's prevalence of deception from being representative of all hotel reviews. Notably, estimates of the prevalence of deception among \emph{negative} reviews might be very different from our estimates, due to the distinct motives of posting deceptive positive vs.~negative reviews. We discuss this further in Section~\ref{sec:conc_and_fut_work}.

\section{Signal Theory} \label{sec:signal_theory}

\begin{table}[t]
\caption{Signal costs associated with six online review communities, sorted approximately from highest signal cost to lowest. \emph{Posting cost} is \emph{High} if users are required to purchase a product before reviewing it, and \emph{Low} otherwise. \emph{Exposure benefit} is \emph{Low}, \emph{Medium}, or \emph{High} based on the number of reviews in the community (see Table~\ref{table:corp_stats}).}
\label{table:comm_factors}
\begin{center}
\begin{tabular}{l|cc}
\hline
\textsc{community} & \textsc{posting cost} & \textsc{exposure benefit} \\
\hline
\textbf{Orbitz} & High & Low \\
\textbf{Priceline} & High & Medium \\
\textbf{Expedia} & High & Medium \\
\textbf{Hotels.com} & High & Medium \\
\textbf{Yelp} & Low & Low \\
\textbf{TripAdvisor} & Low & High \\
\hline
\end{tabular}
\end{center}
\end{table}

\begin{figure*}[t]
\begin{center}
\subfigure[Orbitz]{ \includegraphics[scale=0.62]{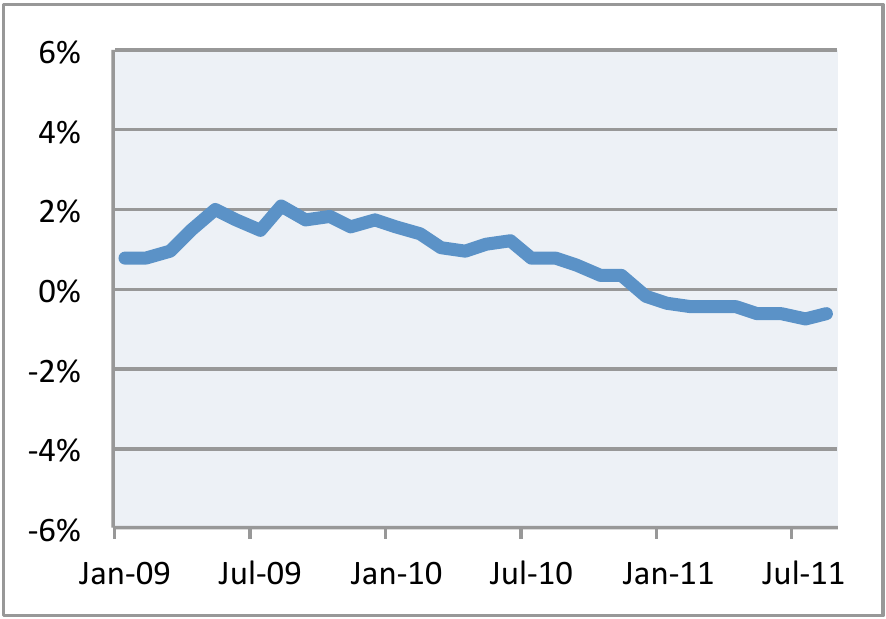} }
\subfigure[Priceline]{ \includegraphics[scale=0.62]{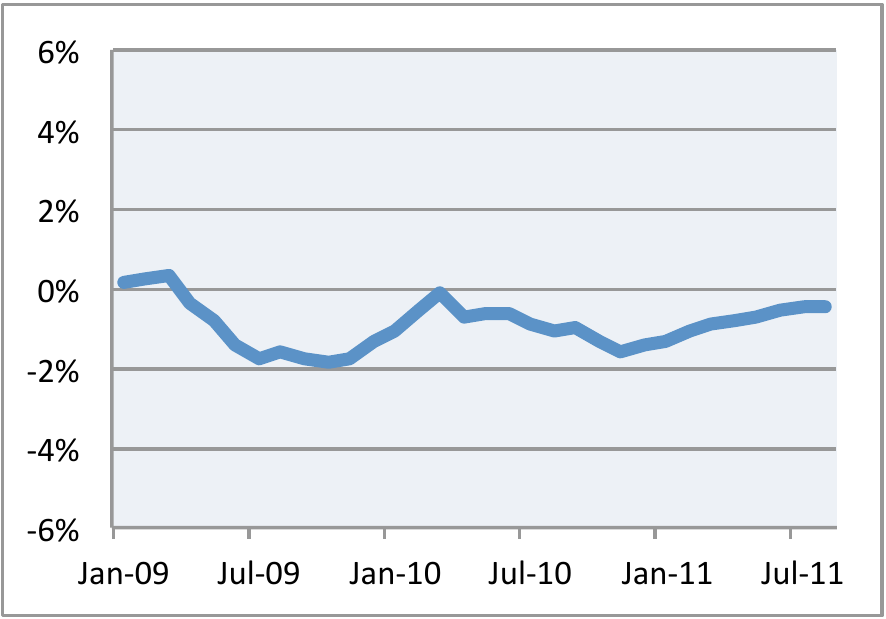} }
\subfigure[Expedia]{ \includegraphics[scale=0.62]{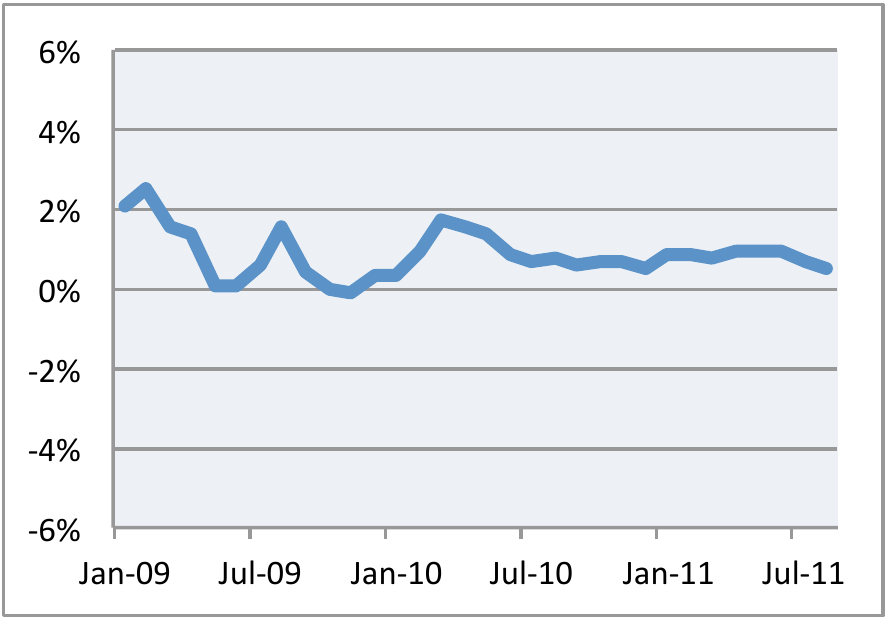} }
\subfigure[Hotels.com]{ \includegraphics[scale=0.62]{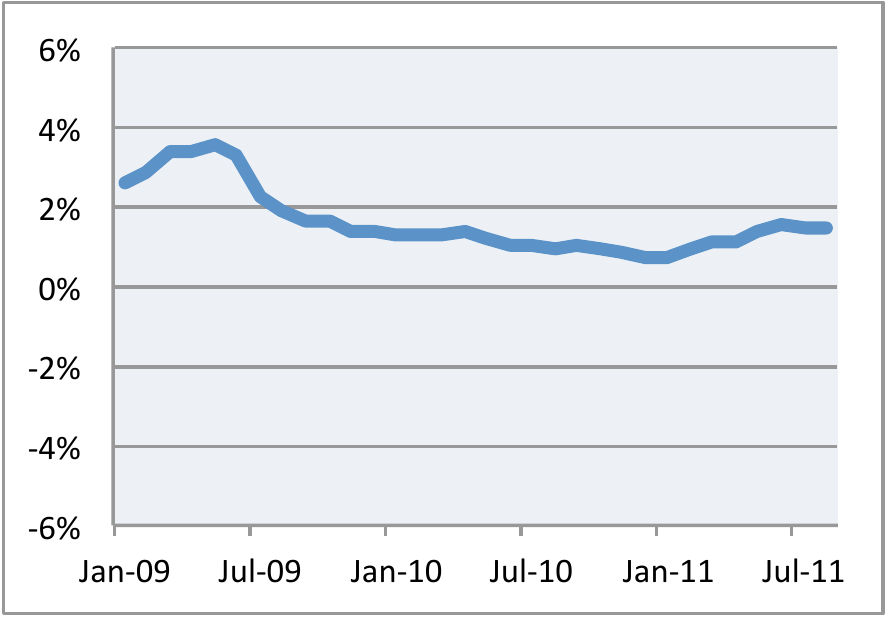} }
\subfigure[Yelp]{ \includegraphics[scale=0.62]{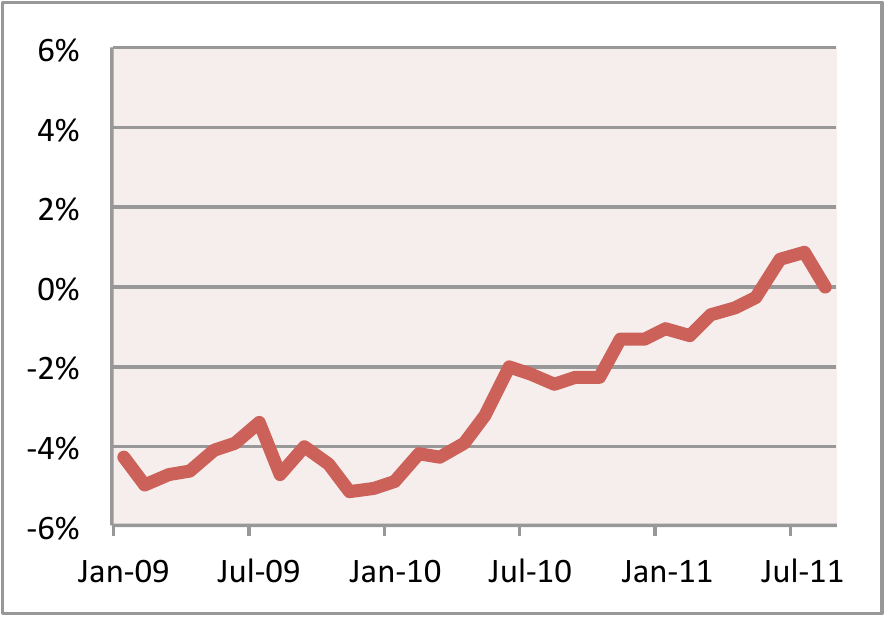} }
\subfigure[TripAdvisor]{ \includegraphics[scale=0.62]{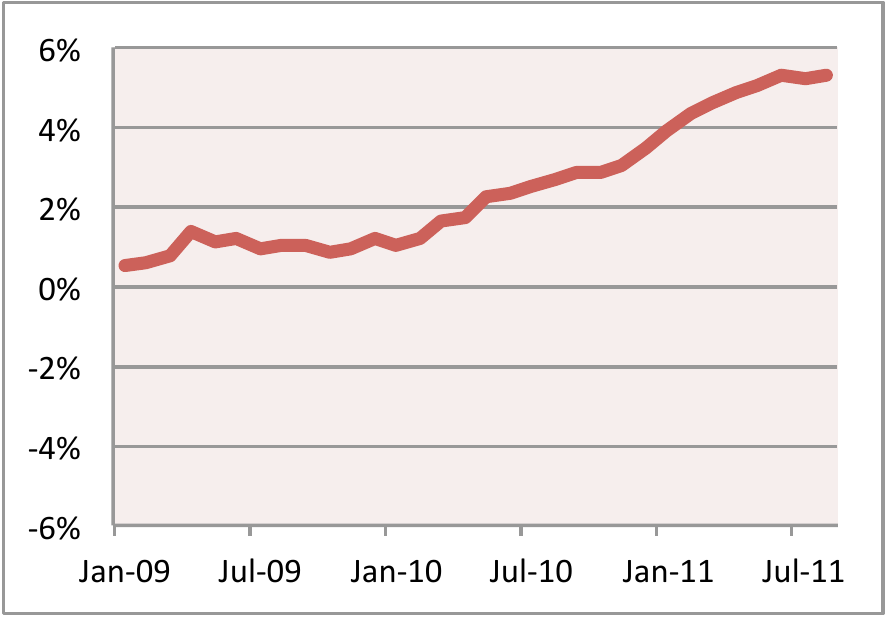} }
\caption{Graph of Na\"ive estimates of deception prevalence versus time, for six online review communities. Blue (a--d) and red (e--f) graphs correspond to high and low posting cost communities, respectively.}
\label{fig:naive}
\end{center}
\end{figure*}

In terms of economic theory, the role of review communities is to reduce the inherent \emph{information asymmetry}~\cite{Spence:73} between buyers and sellers in online marketplaces, by providing buyers with \emph{a priori} knowledge of the underlying quality of the products being sold~\cite{Hu:08}. It follows that if reviews regularly failed to reduce this information asymmetry, or, worse, convey false information, then they would cease to be of value to the user. Given that review communities \emph{are}, in fact, valued by users~\cite{Cone:11}, it seems unlikely that the prevalence of deception among them is large.

Nonetheless, there is widespread concern about the prevalence of deception in online reviews, rightly or wrongly, and further, deceptive reviews can be cause for concern even in small quantities, e.g., if they are \emph{concentrated} in a single review community. We propose that by framing reviews as $signals$---voluntary communications that serve to convey information about the signaler~\cite{Spence:73}, we can reason about the factors underlying deception by manipulating the distinct \emph{signal costs} associated with truthful vs.~deceptive reviews.

Specifically, we claim that for a \emph{positive} review to be posted in a given review community, there must be an incurred \emph{signal cost}, that is \emph{increased} by:
\begin{enumerate}
\item The \textbf{posting cost} for posting the review in a given review community, i.e., whether users are required to purchase a product prior to reviewing it (high cost) or not (low cost). Some sites, for example, allow anyone to post reviews about any hotel, making the review cost effectively zero. Other sites, however, require the purchase of the hotel room before a review can be written, raising the cost from zero to the price of the room.
\end{enumerate}
and \emph{decreased} by:
\begin{enumerate}
\setcounter{enumi}{1}
\item The \textbf{exposure benefit} of posting the review in that review community, i.e., the benefit derived from other users reading the review, which is proportional to the size of the review community's audience. Review sites with more traffic have greater exposure benefit.
\end{enumerate}
Observe that both the posting cost and the exposure benefit depend entirely on the review community. An overview of these factors for each of the six review communities is given in Table~\ref{table:comm_factors}.

Based on the signal cost function just defined, we propose two hypotheses:
\begin{itemize}
\item{\emph{Hypothesis 1}:} Review communities that have low signal costs (low posting requirements, high exposure), e.g., TripAdvisor and Yelp, will have more deception than communities with high signal costs, e.g., Orbitz.
\item{\emph{Hypothesis 2}:} Increasing the signal cost will decrease the prevalence of deception.
\end{itemize}

\section{Experimental Setup} \label{sec:exper_setup}

\begin{figure*}[t]
\begin{center}
\subfigure[Orbitz]{ \includegraphics[scale=0.62]{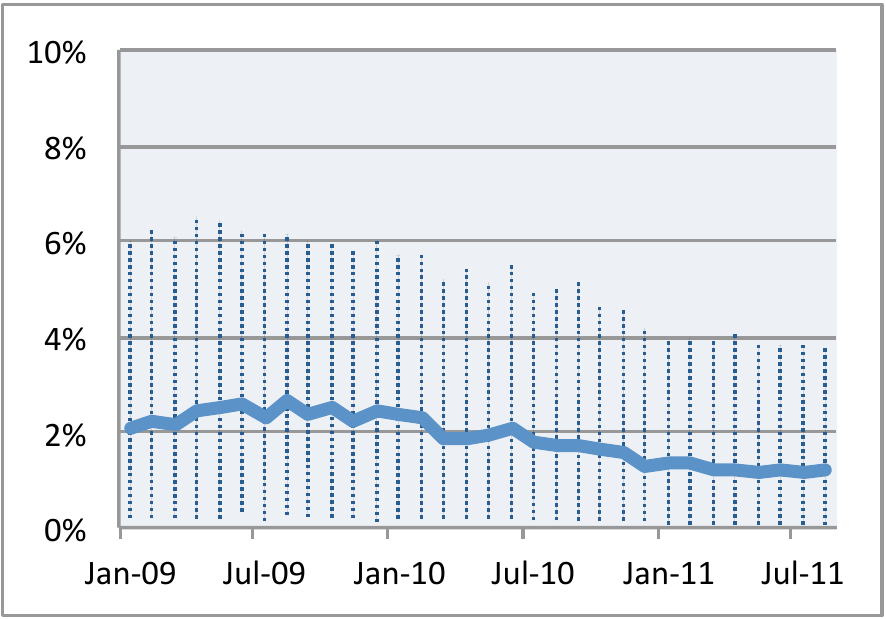} }
\subfigure[Priceline]{ \includegraphics[scale=0.62]{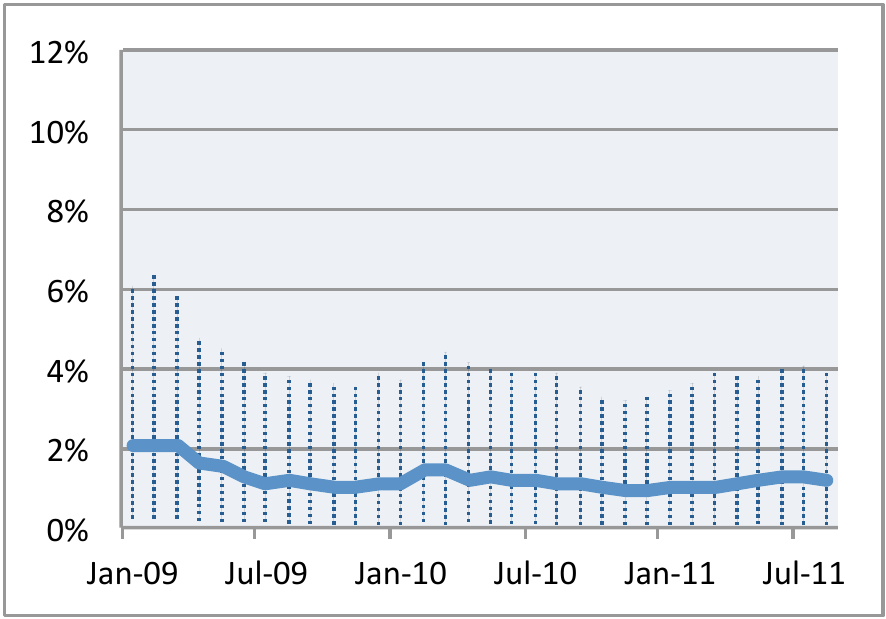} }
\subfigure[Expedia]{ \includegraphics[scale=0.62]{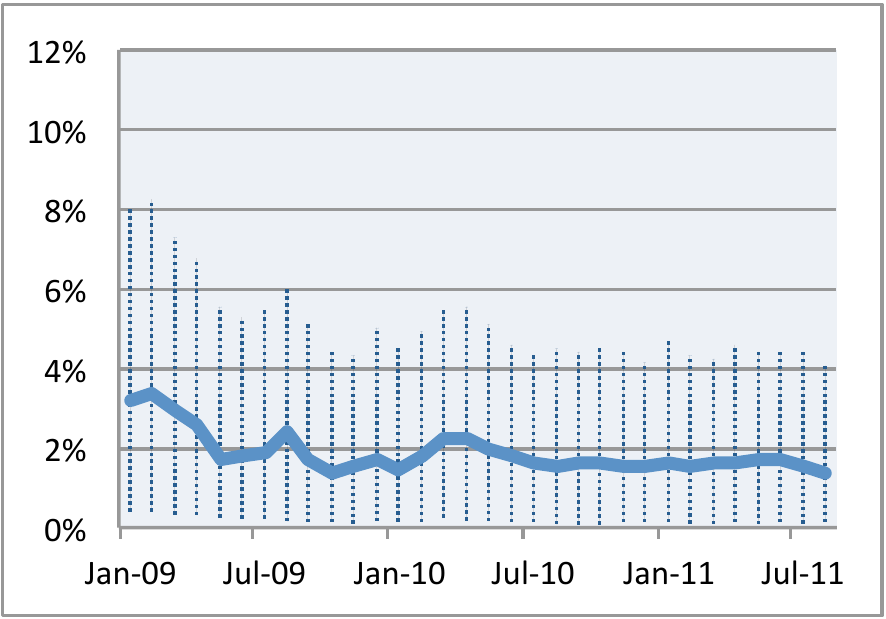} }
\subfigure[Hotels.com]{ \includegraphics[scale=0.62]{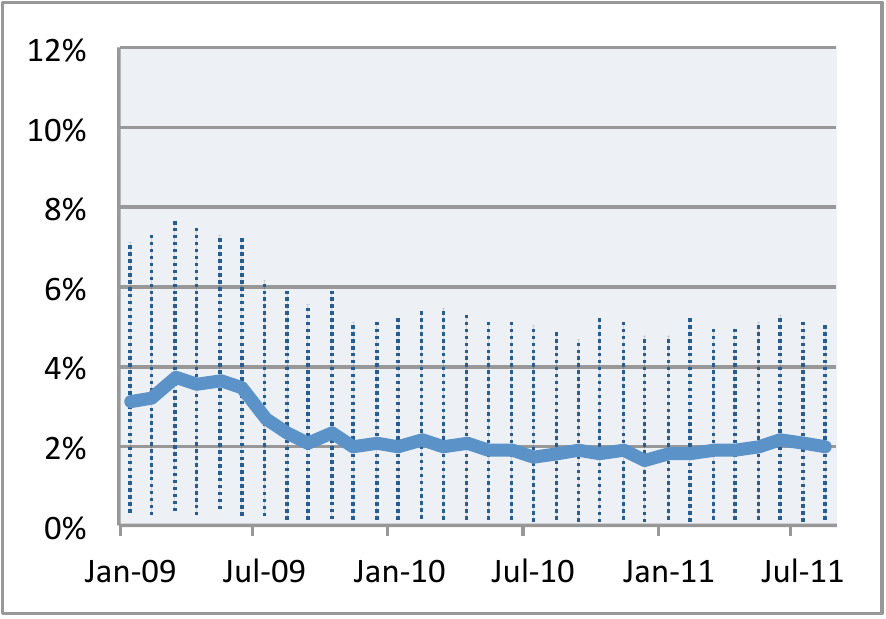} }
\subfigure[Yelp]{ \includegraphics[scale=0.62]{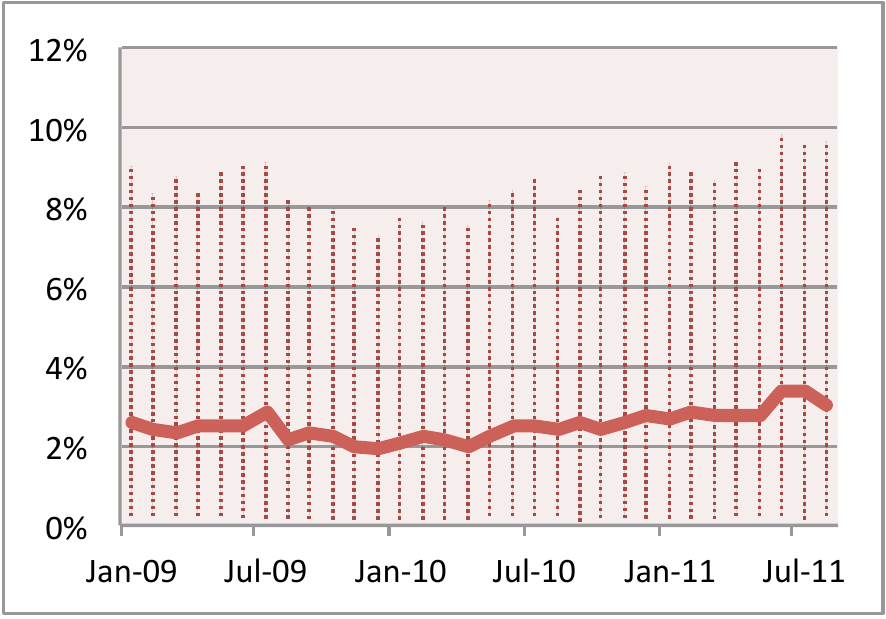} }
\subfigure[TripAdvisor]{ \includegraphics[scale=0.62]{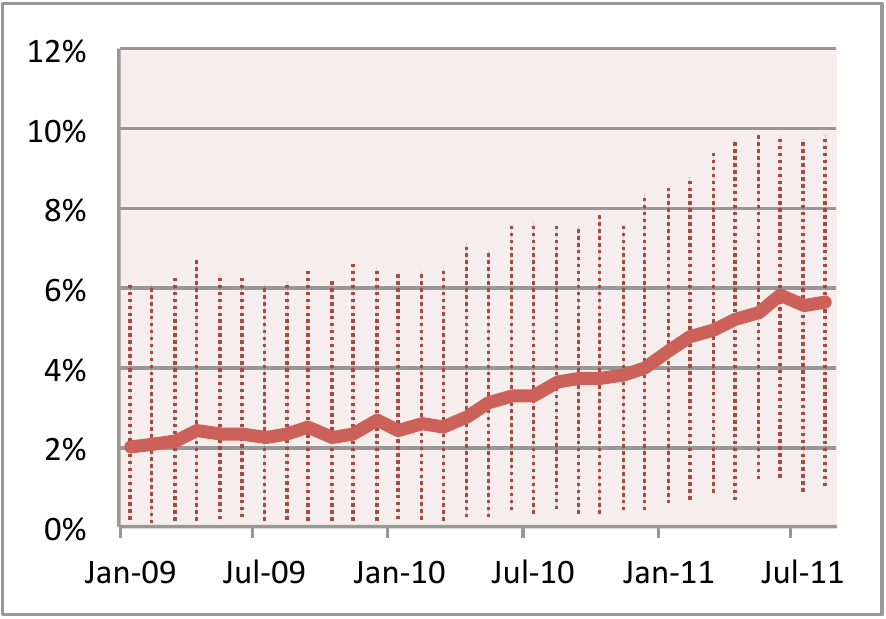} }
\caption{Graph of Bayesian estimates of deception prevalence versus time, for six online review communities. Blue (a--d) and red (e--f) graphs correspond to high and low posting cost communities, respectively. Error bars show Bayesian 95\% credible intervals.}
\label{fig:bayes}
\end{center}
\end{figure*}

The framework described in Section~\ref{sec:frame} is instantiated for the six review communities introduced in Section~\ref{sec:data}. In particular, we first train our SVM deception classifier following the procedure outlined in Section~\ref{sec:dec_class}. An important step when training SVM classifiers is setting the cost parameter, $C$.
We set $C$ using a nested 5-fold cross-validation procedure, and choose the value that gives the best average \emph{balanced accuracy}, defined as $\frac{1}{2} (\text{sensitivity} + \text{specificity})$.

We then estimate the classifier's sensitivity, specificity, and hyperparameters, using the procedure outlined in Section~\ref{sec:class_sens_and_spec} and Appendix~\ref{app:class_sens_and_spec}. Based on those estimates, we then estimate the prevalence of deception among reviews in our test set using the Na\"ive and the Bayesian Prevalence Models. Gibbs sampling for the Bayesian Prevalence Model is performed using Equations~\ref{eqn:gibbs1} and~\ref{eqn:gibbs2} (given in Appendix~\ref{app:bayes_inference}) for 70,000 iterations, with a \emph{burn-in} of 20,000 iterations, and a \emph{sampling lag} of 50. We use an uninformative (uniform) prior for $\pi^*$, i.e., $\boldsymbol{\alpha} = \left< 1, 1 \right>$. Multiple runs are performed to verify the stability of the results.

\section{Results and Discussion} \label{sec:res_and_disc}

\begin{figure*}[t]
\begin{center}
\subfigure[TripAdvisor. All reviews.]{ \includegraphics[scale=0.62]{tripadvisor_excl0} }
\subfigure[TripAdvisor. First-time reviewers excluded.]{ \includegraphics[scale=0.62]{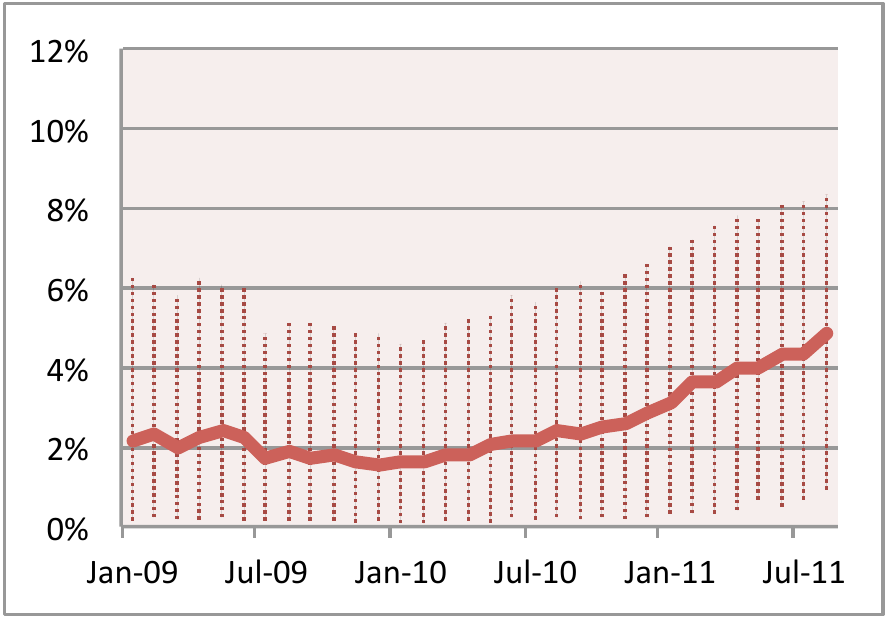} }
\subfigure[TripAdvisor. First-time and second-time reviewers excluded.]{ \includegraphics[scale=0.62]{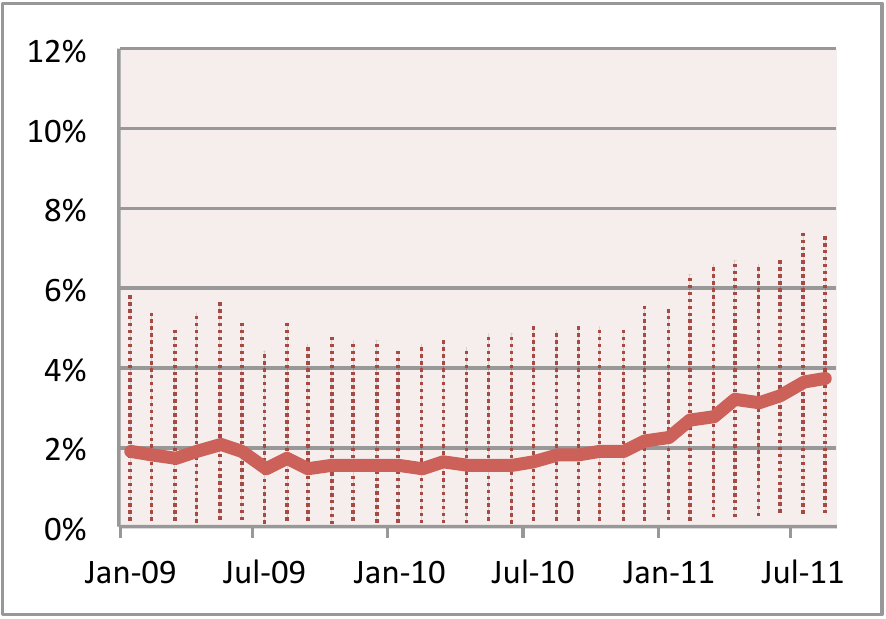} }
\caption{Graph of Bayesian estimates of deception prevalence versus time, for TripAdvisor, with reviews written by new users excluded. Excluding reviews written by first- or second-time reviewers increases the signal cost, and decreases the prevalence of deception.}
\label{fig:signalcost}
\end{center}
\end{figure*}

Estimates of the prevalence of deception for six review communities over time, given by the Na\"ive Prevalence Model, appear in Figure~\ref{fig:naive}. Blue graphs (a--d) correspond to communities with \emph{High} posting cost (see Table~\ref{table:comm_factors}), i.e., communities for which you are required to book a hotel room before posting a review, while red graphs (e--f) correspond to communities with \emph{Low} posting cost, i.e., communities that allow any user to post reviews for any hotel.

In agreement with \emph{Hypothesis 1} (given in Section~\ref{sec:signal_theory}), it is clear from Figure~\ref{fig:naive} that deceptive opinion spam is decreasing or stationary over time for \emph{High} posting cost review communities (blue graphs, a--d). In contrast, review communities that allow any user to post reviews for any hotel, i.e., \emph{Low} posting cost communities (red graphs, e--f), are seeing \emph{growth} in their rate of deceptive opinion spam.

Unfortunately, as discussed in Section~\ref{sec:prev_model_naive}, we observe that the prevalence estimates produced by the Na\"ive Prevalence Model are often negative. This occurs when the rate at which the classifier makes positive predictions is below the classifier's estimated false positive rate, suggesting both that the estimated false positive rate of the classifier is perhaps \emph{overestimated}, and that the classifier's estimated specificity (truthful recall rate, given by $\theta$) is perhaps \emph{underestimated}. We address this further in Section~\ref{sec:assumptions_and_limitations}.

The Bayesian Prevalence Model, on the other hand, encodes the uncertainty in the estimated values of the classifier's sensitivity and specificity through two Beta priors, and in particular their hyperparameters, $\boldsymbol{\beta}$ and $\boldsymbol{\gamma}$. Estimates of the prevalence of deception for the six review communities over time, given by the Bayesian Prevalence Model, appear in Figure~\ref{fig:bayes}. Blue (a--d) and red (e--f) graphs, as before, correspond to communities with \emph{High} and \emph{Low} posting costs, respectively.

In agreement with \emph{Hypothesis 1} (Section~\ref{sec:signal_theory}), we again find that \emph{Low} signal cost communities, e.g., TripAdvisor, seem to contain larger quantities and accelerated growth of deceptive opinion spam when compared to \emph{High} signal cost communities, e.g., Orbitz. Interestingly, communities with a blend of signal costs appear to have medium rates of deception that are neither growing nor declining, e.g., Hotels.com, which has a rate of deception of $\approx 2\%$.

To test \emph{Hypothesis 2}, i.e., that increasing the signal cost will decrease the prevalence of deception, we need to \emph{increase} the signal cost, as we have defined it in Section~\ref{sec:signal_theory}. Thus, it is necessary to either \emph{increase} the posting cost, or \emph{decrease} the exposure benefit. And while we have no control over a community's exposure benefit, we \emph{can} increase the posting cost by, for example, hiding all reviews written by users who have not posted at least two reviews. Essentially, by requiring users to post more than one review in order for their review to be displayed, we are increasing the posting cost and, accordingly, the signal cost as well.

Bayesian Prevalence Model estimates for TripAdvisor for varying signal costs appear in Figure~\ref{fig:signalcost}. In particular, we give the estimated prevalence of deception over time after removing reviews written by first-time review writers, and after removing reviews written by first- or second-time review writers. In agreement with \emph{Hypothesis 2}, we see a clear reduction in the prevalence of deception over time on TripAdvisor after removing these reviews, with rates dropping from $\approx 6\%$, to $\approx 5\%$, and finally to $\approx 4\%$, suggesting that an increased signal cost may indeed help to reduce the prevalence of deception in online review communities.

\subsection{Assumptions and Limitations} \label{sec:assumptions_and_limitations}

In this work we have made a number of assumptions, a few of which we will now highlight and discuss.

First, we note that our unlabeled test set, $\mathcal{D}^{\text{test}}$, overlaps with our labeled truthful training set, $\mathcal{D}^{\text{train}}$. Consequently, we will \emph{underestimate} the prevalence of deception, because the overlapping reviews will be more likely to be classified at test time as truthful, having been seen in training as being truthful. Excluding these overlapping reviews from the test set results in \emph{overestimating} the prevalence of deception, based on the hypothesis that the overlapping reviews, chosen from the 20 most highly-reviewed Chicago hotels, are more likely to be truthful to begin with.

Second, we observe that our development set, $\mathcal{D}^{\text{dev}}$, containing \emph{labeled truthful} reviews, is not gold-standard. Unfortunately, while it is necessary to obtain a uniform sample of reviews in order to fairly estimate the classifier's truthful recall rate (\emph{specificity}), such review samples are inherently unlabeled. This can be problematic if the underlying rate of deception is high among the reviews from which the development set is sampled, because the specificity will then be \emph{underestimated}. Indeed, our Na\"ive Prevalence Model regularly produces negative estimates, suggesting that the estimated classifier specificity may indeed be underestimated, possibly due to deceptive reviews in the development set.

Third, our proposal for increasing the signal cost, by hiding reviews written by first- or second-time reviewers, is not ideal. While our results confirm that hiding these reviews will cause an immediate reduction in deception prevalence, the increase in signal cost might be insufficient to discourage new deception, once deceivers become aware of the increased posting requirements.

Fourth, in this work we have only considered a limited version of the deception prevalence problem. In particular, we have only considered positive Chicago hotel reviews, and our classifier is trained on deceptive reviews coming only from Amazon Mechanical Turk. Both negative reviews as well as deceptive reviews obtained by other means are likely to be different in character than the data used in this study.

\subsection{Implications for Psychological Research} \label{sec:implications}

The current research also represents a novel approach to a long-standing and ongoing debate around deception prevalence in the psychological literature. In one of the first large-scale studies looking at how often people lie in everyday communication, DePaulo et al.~\cite{DePaulo:96} used a diary method to calculate the average number of lies told per day. At the end of seven days participants told approximately one to two lies per day, with more recent studies replicating this general finding~\cite{Hancock:04}, suggesting that deception is frequent in human communication. More recently, Serota et al.~\cite{Serota:10}
conducted a large scale representative survey of Americans asking participants how often they lied in the last 24 hours. While they found the same average deception rate as previous research (approximately 1.65 lies per day), they discovered that the data was heavily skewed, with 60 percent of the participants reporting no lies at all. They concluded that rather than deception prevalence being spread evenly across the population, there are instead a few prolific liars. Unfortunately, both sides of this debate have relied solely on self-report data.

The current approach offers a novel method for assessing deception prevalence that does not require self-report, but can provide insight into the prevalence of deception in human communication more generally. At the same time, the question raised by the psychological research also mirrors an important point regarding the prevalence of deception in online reviews: are a few deceptive reviews posted by many people, or are there many deceptive reviews told by only a few? That is, do some hotels have many fake reviews while others are primarily honest? Or, is there a little bit of cheating by most hotels? This kind of individualized modeling represents an important next step in this line of research.

\section{Conclusion} \label{sec:conc_and_fut_work}

In this work, we have presented a general framework for estimating the prevalence of deception in online review communities, based on the output of a noisy deception classifier. Using this framework, we have explored the prevalence of deception among positive reviews in six popular online review communities, and provided the first empirical study of the magnitude, and influencing factors of deceptive opinion spam.

We have additionally proposed a theoretical model of online reviews as a signal to a product's true (unknown) quality, based on economic signaling theory. Specifically, we have defined the signal cost of positive online reviews as a function of the posting costs and exposure benefits of the review community in which it is posted. Based on this theory, we have further suggested two hypotheses, both of which are supported by our findings. In particular, we find first that review communities with low signal costs (low posting requirements, high exposure) have \emph{more deception} than communities with comparatively higher signal costs. Second, we find that by increasing the signal cost of a review community, e.g., by excluding reviews written by first- or second-time reviewers, we can effectively reduce both the prevalence and the growth rate of deception in that community.

Future work might explore other methods for manipulating the signal costs associated with posting online reviews, and the corresponding effects on deception prevalence. For example, some sites, such as Angie's List (\url{http://www.angieslist.com/}), charge a monthly access fee in order to browse or post reviews, and future work might study the effectiveness of such techniques at deterring deception.

\section{Acknowledgments}

This work was supported in part by National Science Foundation Grant NSCC-0904913, and the Jack Kent Cooke Foundation. We also thank, alphabetically, Cristian Danescu-Niculescu-Mizil, Lillian Lee, Bin Lu, Karthik Raman, Lu Wang, and Ainur Yessenalina, as well as members of the Cornell NLP seminar group and the WWW reviewers for their insightful comments, suggestions and advice on various aspects of this work.

\bibliographystyle{abbrv}

\begin{thebibliography}{10}

\bibitem{Bond:06}
C.~Bond and B.~DePaulo.
\newblock {Accuracy of deception judgments}.
\newblock {\em Personality and Social Psychology Review}, 10(3):214, 2006.

\bibitem{CC01a}
C.-C. Chang and C.-J. Lin.
\newblock {LIBSVM}: A library for support vector machines.
\newblock {\em ACM Transactions on Intelligent Systems and Technology},
  2:27:1--27:27, 2011.
\newblock Software available at \url{http://www.csie.ntu.edu.tw/~cjlin/libsvm}.

\bibitem{Cone:11}
Cone.
\newblock {2011 Online Influence Trend Tracker}.
\newblock Online:
  \url{http://www.coneinc.com/negative-reviews-online-reverse-purchase-decisions},
  August 2011.

\bibitem{DePaulo:96}
B.~DePaulo, D.~Kashy, S.~Kirkendol, M.~Wyer, and J.~Epstein.
\newblock Lying in everyday life.
\newblock {\em Journal of personality and social psychology}, 70(5):979, 1996.

\bibitem{Hancock:09:DMCC}
J.~Hancock.
\newblock {Digital Deception: The Practice of Lying in the Digital Age}.
\newblock {\em Deception: Methods, Contexts and Consequences}, pages 109--120,
  2009.

\bibitem{Hancock:04}
J.~Hancock, J.~Thom-Santelli, and T.~Ritchie.
\newblock Deception and design: The impact of communication technology on lying
  behavior.
\newblock In {\em Proceedings of the SIGCHI conference on Human factors in
  computing systems}, pages 129--134. ACM, 2004.

\bibitem{Hu:08}
N.~Hu, L.~Liu, and J.~Zhang.
\newblock Do online reviews affect product sales? The role of reviewer
  characteristics and temporal effects.
\newblock {\em Information Technology and Management}, 9(3):201--214, 2008.

\bibitem{Jindal:08}
N.~Jindal and B.~Liu.
\newblock {Opinion spam and analysis}.
\newblock In {\em Proceedings of the international conference on Web search and
  web data mining}, pages 219--230. ACM, 2008.

\bibitem{Johnson:01}
W.~Johnson, J.~Gastwirth, and L.~Pearson.
\newblock Screening without a ``gold standard": The Hui-Walter paradigm
  revisited.
\newblock {\em American Journal of Epidemiology}, 153(9):921, 2001.

\bibitem{Joseph:95}
L.~Joseph, T.~Gyorkos, and L.~Coupal.
\newblock Bayesian estimation of disease prevalence and the parameters of
  diagnostic tests in the absence of a gold standard.
\newblock {\em American Journal of Epidemiology}, 141(3):263, 1995.

\bibitem{Lim:10}
E.~Lim, V.~Nguyen, N.~Jindal, B.~Liu, and H.~Lauw.
\newblock {Detecting product review spammers using rating behaviors}.
\newblock In {\em Proceedings of the 19th ACM international conference on
  Information and knowledge management}, pages 939--948. ACM, 2010.

\bibitem{CNET:09}
D.~Meyer.
\newblock Fake reviews prompt belkin apology.
\newblock \url{http://news.cnet.com/8301-1001_3-10145399-92.html}, Jan. 2009.

\bibitem{Mihalcea:09}
R.~Mihalcea and C.~Strapparava.
\newblock {The lie detector: Explorations in the automatic recognition of
  deceptive language}.
\newblock In {\em Proceedings of the ACL-IJCNLP 2009 Conference Short Papers},
  pages 309--312. Association for Computational Linguistics, 2009.

\bibitem{NYTimes:09}
C.~Miller.
\newblock Company settles case of reviews it faked.
\newblock
  \url{http://www.nytimes.com/2009/07/15/technology/internet/15lift.html}, July
  2009.

\bibitem{Ott:11}
M.~Ott, Y.~Choi, C.~Cardie, and J.~Hancock.
\newblock Finding deceptive opinion spam by any stretch of the imagination.
\newblock In {\em Proceedings of the 49th Annual Meeting of the Association for
  Computational Linguistics: Human Language Technologies-Volume 1}, pages
  309--319. Association for Computational Linguistics, 2011.

\bibitem{Guardian:11}
B.~Page.
\newblock Amazon withdraws ebook explaining how to manipulate its sales
  rankings.
\newblock
  \url{http://www.guardian.co.uk/books/2011/jan/05/amazon-ebook-manipulate-kindle-rankings},
  Jan. 2011.

\bibitem{Serota:10}
K.~Serota, T.~Levine, and F.~Boster.
\newblock The prevalence of lying in America: Three studies of self-reported
  lies.
\newblock {\em Human Communication Research}, 36(1):2--25, 2010.

\bibitem{Spence:73}
M.~Spence.
\newblock Job market signaling.
\newblock {\em The quarterly journal of Economics}, 87(3):355, 1973.

\bibitem{NYTimes:11}
D.~Streitfeld.
\newblock In a race to out-rave, 5-star web reviews go for \$5.
\newblock
  \url{http://www.nytimes.com/2011/08/20/technology/finding-fake-reviews-online.html},
  Aug. 2011.

\bibitem{NYTimes:12}
D.~Streitfeld.
\newblock For \$2 a star, an online retailer gets 5-star product reviews.
\newblock
  \url{http://www.nytimes.com/2012/01/27/technology/for-2-a-star-a-retailer-gets-5-star-reviews.html},
  Jan. 2012.

\bibitem{Guardian:10}
A.~Topping.
\newblock Historian Orlando Figes agrees to pay damages for fake reviews.
\newblock
  \url{http://www.guardian.co.uk/books/2010/jul/16/orlando-figes-fake-amazon-reviews},
  July 2010.

\bibitem{Vrij:08}
A.~Vrij.
\newblock {\em {Detecting lies and deceit: Pitfalls and opportunities}}.
\newblock Wiley-Interscience, 2008.

\bibitem{Yoo:09}
K.~Yoo and U.~Gretzel.
\newblock {Comparison of Deceptive and Truthful Travel Reviews}.
\newblock {\em Information and Communication Technologies in Tourism 2009},
  pages 37--47, 2009.

\end{thebibliography}

\appendix

\section{Gibbs Sampler for Bayesian Prevalence Model} \label{app:bayes_inference}

Gibbs sampling of the Bayesian Prevalence Model, introduced in Section~\ref{sec:prev_model_bayes}, is performed according to the following conditional distributions:
\begin{multline}
\Pr(y_i = 1 \given f(\mathbf{x}), \mathbf{y}^{(-i)}; \boldsymbol{\alpha}, \boldsymbol{\beta}, \boldsymbol{\gamma}) \\
\propto (\boldsymbol{\alpha}_1 + N_1^{(-i)}) \cdot \frac{\boldsymbol{\beta}_{f(\mathbf{x}_i)} + X_{f(\mathbf{x}_i)}^{(-i)}}{\sum \boldsymbol{\beta} + N_1^{(-i)}}, \label{eqn:gibbs1}
\end{multline}
and,
\begin{multline}
\Pr(y_i = 0 \given f(\mathbf{x}), \mathbf{y}^{(-i)}; \boldsymbol{\alpha}, \boldsymbol{\beta}, \boldsymbol{\gamma}) \\
\propto (\boldsymbol{\alpha}_0 + N_0^{(-i)}) \cdot \frac{\boldsymbol{\gamma}_{1 - f(\mathbf{x}_i)} + Y_{f(\mathbf{x}_i)}^{(-i)}}{\sum \boldsymbol{\gamma} + N_0^{(-i)}}, \label{eqn:gibbs2}
\end{multline}
where,
\begin{align*}
X_k^{(-i)} &= \sum_{j \neq i} \sigma[y_j = 1] \cdot \sigma[f(\mathbf{x}_j) = k], \\
Y_k^{(-i)} &= \sum_{j \neq i} \sigma[y_j = 0] \cdot \sigma[f(\mathbf{x}_j) = k], \\
N_1^{(-i)} &= X_0^{(-i)} + X_1^{(-i)}, \\
N_0^{(-i)} &= Y_0^{(-i)} + Y_1^{(-i)}.
\end{align*}
After sampling, we reconstruct the collapsed variables to yield the Bayesian Prevalence Model estimate of the prevalence of deception:
\begin{align}
\pi_{\textsc{bayes}} &= \frac{\boldsymbol{\alpha}_1 + N_1}{\sum \boldsymbol{\alpha} + N^{\text{test}}}. \label{eqn:bayes_prev_est}
\end{align}
Estimates of the classifier's sensitivity and specificity are similarly given by:
\begin{align}
\eta_{\textsc{bayes}} &= \frac{\boldsymbol{\beta}_1 + X_1}{\sum \boldsymbol{\beta} + N_1}, \\
\theta_{\textsc{bayes}} &= \frac{\boldsymbol{\gamma}_1 + Y_0}{\sum \boldsymbol{\gamma} + N_0}.
\end{align}

\balancecolumns

\section{Estimating Classifier Sensitivity and Specificity} \label{app:class_sens_and_spec}

We estimate the sensitivity and specificity of our deception classifier via the following procedure:

\begin{enumerate}
\item Assume given a labeled training set, $\mathcal{D}^{\text{train}}$, containing $N^{\text{train}}$ reviews of $n$ hotels. Also assume given a development set, $\mathcal{D}^{\text{dev}}$, containing labeled truthful reviews.
\item Split $\mathcal{D}^{\text{train}}$ into $n$ folds, $\mathcal{D}^{\text{train}}_1, \dots, \mathcal{D}^{\text{train}}_n$, of sizes given by, $N^{\text{train}}_1, \dots, N^{\text{train}}_n$, respectively, such that $\mathcal{D}^{\text{train}}_j$ contains all (and only) reviews of hotel $j$. Let $\mathcal{D}^{\text{train}}_{(-j)}$ contain all reviews \emph{except} those of hotel $j$.
\item Then, for each hotel $j$: \begin{enumerate}
  \item Train a classifier, $f_j$, from reviews in $\mathcal{D}^{\text{train}}_{(-j)}$, and use it to classify reviews in $\mathcal{D}^{\text{train}}_j$.
  \item Let $|TP|_j$ correspond to the observed number of true positives, i.e.:
    \begin{align}
    |TP|_j = \sum_{(\mathbf{x}, y) \in \mathcal{D}^{\text{train}}_j} \sigma[y = 1] \cdot \sigma[f_j(\mathbf{x}) = 1].
    \end{align}
  \item Similarly, let $|FN|_j$ correspond to the observed number of false negatives.
  \end{enumerate}
\item Calculate the aggregate number of true positives ($|TP|$) and false negatives ($|FN|$), and compute the sensitivity (deceptive recall) as:
  \begin{align}
  \eta = \frac{|TP|}{|TP| + |FN|}.
  \end{align}
\item Train a classifier using \emph{all} reviews in $\mathcal{D}^{\text{train}}$, and use it to classify reviews in $\mathcal{D}^{\text{dev}}$.
\item Let the resulting number of true negative and false positive predictions in $\mathcal{D}^{\text{dev}}$ be given by $|TN|_{dev}$ and $|FP|_{dev}$, respectively, and compute the specificity (truthful recall) as:
  \begin{align}
  \theta = \frac{|TN|_{dev}}{|TN|_{dev} + |FP|_{dev}}.
  \end{align}
\end{enumerate}

For the Bayesian Prevalence Model, we observe that the posterior distribution of a variable with an uninformative (uniform) $\text{Beta}$ prior, after observing $a$ successes and $b$ failures, is just $\text{Beta}(a + 1, b + 1)$, i.e., $a$ and $b$ are \emph{pseudo counts}. Based on this observation, we set the hyperparameters $\boldsymbol{\beta}$ and $\boldsymbol{\gamma}$, corresponding to the classifier's sensitivity (deceptive recall) and specificity (truthful recall), respectively, to:
\begin{align*}
\boldsymbol{\beta} &= \left< |FN| + 1 , |TP| + 1 \right>,\\
\boldsymbol{\gamma} &= \left< |FP|_{dev} + 1, |TN|_{dev} + 1 \right>.
\end{align*}

\end{document}